\documentclass{article}
\usepackage{lscape,epsfig}
\usepackage{graphicx}
\usepackage{amssymb}
\usepackage{amsmath}
\usepackage{natbib}
\textheight=22cm \DeclareSymbolFont{ppa}{OT1}{ppl}{m}{it}
\DeclareMathSymbol{\vv}{\mathalpha}{ppa}{'166}

minus 2.3mu \thickmuskip = 2.6mu plus 2mu minus 2.6mu

\begin{document}

\newcommand{\dd}{\,{\rm d}}
\newcommand{\ie}{{\it i.e.},\,}
\newcommand{\etal}{{\it $et$ $al$.\ }}
\newcommand{\eg}{{\it e.g.},\,}
\newcommand{\cf}{{\it cf.\ }}
\newcommand{\vs}{{\it vs.\ }}
\newcommand{\zdot}{\makebox[0pt][l]{.}}
\newcommand{\up}[1]{\ifmmode^{\rm #1}\else$^{\rm #1}$\fi}
\newcommand{\dn}[1]{\ifmmode_{\rm #1}\else$_{\rm #1}$\fi}
\newcommand{\upd}{\up{d}}
\newcommand{\uph}{\up{h}}
\newcommand{\upm}{\up{m}}
\newcommand{\ups}{\up{s}}
\newcommand{\arcd}{\ifmmode^{\circ}\else$^{\circ}$\fi}
\newcommand{\arcm}{\ifmmode{'}\else$'$\fi}
\newcommand{\arcs}{\ifmmode{''}\else$''$\fi}
\newcommand{\MS}{{\rm M}\ifmmode_{\odot}\else$_{\odot}$\fi}
\newcommand{\RS}{{\rm R}\ifmmode_{\odot}\else$_{\odot}$\fi}
\newcommand{\LS}{{\rm L}\ifmmode_{\odot}\else$_{\odot}$\fi}

\newcommand{\Abstract}[2]{{\footnotesize\begin{center}ABSTRACT\end{center}
\vspace{1mm}\par#1\par \noindent {~}{\it #2}}}

\newcommand{\TabCap}[2]{\begin{center}\parbox[t]{#1}{\begin{center}
  \small {\spaceskip 2pt plus 1pt minus 1pt T a b l e}
  \refstepcounter{table}\thetable \\[2mm]
  \footnotesize #2 \end{center}}\end{center}}

\newcommand{\TableSep}[2]{\begin{table}[p]\vspace{#1}
\TabCap{#2}\end{table}}

\newcommand{\FigCap}[1]{\footnotesize\par\noindent Fig.\  %
  \refstepcounter{figure}\thefigure. #1\par}

\newcommand{\TableFont}{\footnotesize}
\newcommand{\TableFontIt}{\ttit}
\newcommand{\SetTableFont}[1]{\renewcommand{\TableFont}{#1}}

\newcommand{\MakeTable}[4]{\begin{table}[htb]\TabCap{#2}{#3}
  \begin{center} \TableFont \begin{tabular}{#1} #4
  \end{tabular}\end{center}\end{table}}

\newcommand{\MakeTableSep}[4]{\begin{table}[p]\TabCap{#2}{#3}
  \begin{center} \TableFont \begin{tabular}{#1} #4
  \end{tabular}\end{center}\end{table}}

\newenvironment{references}%
{ \footnotesize \frenchspacing
\renewcommand{\thesection}{}
\renewcommand{\in}{{\rm in }}
\renewcommand{\AA}{Astron.\ Astrophys.}
\newcommand{\AAS}{Astron.~Astrophys.~Suppl.~Ser.}
\newcommand{\ApJ}{Astrophys.\ J.}
\newcommand{\ApJS}{Astrophys.\ J.~Suppl.~Ser.}
\newcommand{\ApJL}{Astrophys.\ J.~Letters}
\newcommand{\AJ}{Astron.\ J.}
\newcommand{\IBVS}{IBVS}
\newcommand{\PASP}{P.A.S.P.}
\newcommand{\Acta}{Acta Astron.}
\newcommand{\MNRAS}{MNRAS}
\renewcommand{\and}{{\rm and }}
\section{{\rm REFERENCES}}
\sloppy \hyphenpenalty10000
\begin{list}{}{\leftmargin1cm\listparindent-1cm
\itemindent\listparindent\parsep0pt\itemsep0pt}}%
{\end{list}\vspace{2mm}}

\def\TYLDA{~}
\newlength{\DW}
\settowidth{\DW}{0}
\newcommand{\dw}{\hspace{\DW}}

\newcommand{\refitem}[5]{\item[]{#1} #2%
\def\REFARG{#3}\ifx\REFARG\TYLDA\else, {\it#3}\fi
\def\REFARG{#4}\ifx\REFARG\TYLDA\else, {\bf#4}\fi
\def\REFARG{#5}\ifx\REFARG\TYLDA\else, {#5}\fi.}

\newcommand{\Section}[1]{\section{#1}}
\newcommand{\Subsection}[1]{\subsection{#1}}
\newcommand{\Acknow}[1]{\par\vspace{5mm}{\bf Acknowledgements.} #1}
\pagestyle{myheadings}

\newfont{\bb}{ptmbi8t at 12pt}
\newcommand{\xrule}{\rule{0pt}{2.5ex}}
\newcommand{\xxrule}{\rule[-1.8ex]{0pt}{4.5ex}}
\def\thefootnote{\fnsymbol{footnote}}

\begin{center}
 {\Large\bf
  A proper motion study of the globular clusters 
  M4, M12, M22, NGC~3201, NGC~6362 and NGC~6752 
  \footnote {Based 
  on data obtained at the Las Campanas Observatory.}}
  \vskip1cm
  {\large
        K.~~Z~l~o~c~z~e~w~s~k~i$^1$,
      ~~J.~~K~a~l~u~z~n~y$^1$,
      ~~M.~~R~o~z~y~c~z~k~a$^1$,
      ~~W.~~K~r~z~e~m~i~n~s~k~i$^1$,
      ~~B.~~M~a~z~u~r$^1$,
      ~~and~~I.~B.~~T~h~o~m~p~s~o~n$^2$\\
  }
  \vskip3mm {
  $^1$Nicolaus Copernicus Astronomical Center, ul. Bartycka 18, 00-716 Warsaw, Poland\\
     e-mail: (kzlocz, jka, mnr, wk, batka)@camk.edu.pl\\
  $^2$The Observatories of the Carnegie Institution of Washington, \\813 Santa Barbara
      Street, Pasadena, CA 91101, USA\\
     e-mail: ian@obs.carnegiescience.edu}
\end{center}
\Abstract 
{ We derive relative proper motions 
of stars in the fields of globular clusters M4, M12, M22, NGC~3201, NGC~6362 
and NGC~6752 based on a uniform data set collected between 1997 and 2008.
We assign a membership class for each star with a measured proper motion, and show 
that these membership classes can be successfully used to eliminate field stars from 
color-magnitude diagrams of the clusters. They also allow for the efficient selection 
of rare objects such as blue/yellow/red stragglers and stars from the asymptotic giant
branch. Tables with proper motions and photometry 
of over 87000 stars are made publicly available via the Internet. 
 }

{\bf Key words:} {\it globular clusters: individual: M4, M12, M22, NGC~3201,
                 NGC~6362, NGC~6752 -- astrometry -- blue stragglers
                 }
\section {Introduction} \label{sect:intro}
Globular clusters (GCs) offer an excellent opportunity to study the chemical
evolution of the universe, to test the theory of stellar evolution, and
to assess the quality and accuracy of numerical codes written to follow
the dynamical evolution of multi-body systems. To fully exploit this
opportunity, cluster members must be discerned from field stars, which can be
achieved e.g. by a proper motion (PM) study. As emphasized by Bellini, Anderson
and van der Marel (2012), such studies impose strong constraints on structure,
dynamics and evolution of GCs. They also allow the identification of cluster members
occupying unusual locations in the H-R diagram (e.g. blue, yellow and
red stragglers), and, when supplemented with additional data, a measurement of
absolute motions and geometric distances of GCs.

The first PM studies of GCs date from the 1970's. However the statistical uncertainties
were often larger than the internal velocity dispersion \citep[][and references
therein]{cud80}. The breakthrough came with the advent of CCD detectors and the
launch of the Hubble Space Telescope. Extensive HST-studies of numerous GCs have
been undertaken, most of which are not yet finished \citep{bel12}. To our knowledge,
just two large catalogs of PMs derived from Hubble data are accessible online as
of today:  47~Tuc \citep{mcl06} and $\omega$~Cen \citep{and10}. A valuable
extension of the latter is the catalog compiled by \citet{bel09} using ESO archive
CCD data.

As \citet{bel09} show, ground-based CCD observations with temporal
baselines of 4 years are sufficient to reliably separate members of nearby
clusters from field stars. Similar conclusions have been reached by \citet{and06}
for the globular clusters M4 and NGC~6397, \citet{yad08} for the open cluster M67,
and \citet{mon09} for the open cluster NGC~6253.

More recently Zloczewski, Kaluzny \& Thompson (2011) published a PM catalog for M55
based on CCD observations performed at
the Las Campanas Observatory (LCO) between 1997 and 2008. In the present paper we
derive relative PMs for another six GCs: M4, M12, M22, NGC~3201, NGC~6362 and
NGC~6752. In Sec. 2 we describe the observational data and methods used to prepare
the CCD images for the analysis. The procedures employed for the determination of
relative PMs of individual stars are discussed in Sec. 3. Color-magnitude diagrams 
(CMD) of individual clusters, cleaned from field stars, are presented in Sec. 4
followed by a brief Summary.
\section {Data selection and preparation}
\label{sect:data}
The images analyzed in this paper are a part of the data collected
between 1997 and 2008 within the CASE project
\citep[Cluster AgeS Experiment; see][]{jka05a}.
All observations were made with the 2.5-m du Pont telescope at LCO
using the same detector and the same set of $V$ and $I$
filters. We used the TEK5 CCD camera with a field of view of
530$\times$530 arcsec$^2$ and a scale of 0.259 arcsec/pixel. We observed two
fields for M4, two for NGC~3201 and one for each of 
the remaining four clusters. Equatorial coordinates of the centers of 
the analyzed fields are listed in Table \ref{tab:fields}.
\begin{table}
\caption{Equatorial coordinates of field centers.
\label{tab:fields}}
\begin{tabular}{@{}lcc}
\hline
Field & $\alpha_{2000}$ & $\delta_{2000}$ \\
 & [deg] & [deg] \\ 
\hline
M4-F1        & 245.89492 &  -26.52768\\
M4-F2        & 245.90846 &  -26.58710\\
M12          & 251.81906 &   -1.95024\\
M22          & 279.10275 &  -23.90205\\
NGC~3201-F1  & 154.40702 &  -46.40682\\
NGC~3201-F2  & 154.37700 &  -46.39437\\
NGC~6362     & 262.98281 &  -67.04611\\
NGC~6752     & 287.72575 &  -59.97099\\
\hline
\end{tabular}
\end{table}
For each cluster and each observing run we selected a collection of
$V$-frames obtained at an air-mass less than 1.1 and a seeing better
than 1.1 arcsec, henceforth referred to as a {\em data set}. Frames obtained
through cirrus, as well as those with a bright background, were rejected.
Altogether we collected 42, 24, 15, 33, 27 and 22 data sets
for M4, M12, M22, NGC~3201, NGC~6362 and NGC~6752, respectively. For each data set we
constructed an {\em averaged frame} with a high signal-to-noise ratio using
the Difference Image Analysis PL (DIAPL) package\footnote{Freely available
at http://users.camk.edu.pl/pych/DIAPL/} developed by Wojtek
Pych. The construction procedure consisted of the following steps:
\begin{enumerate}
 \item Find the frame with the best seeing (henceforth {\em the reference frame}).
 \item Transform the remaining frames to the coordinates of the reference frame
       (bicubic spline interpolation was used).
 \item Find the point spread function (PSF) of each frame and transform it to
       match that of the reference frame.
 \item Stack the transformed images.
\end{enumerate}
To reduce the effects of PSF variability, the reference frame as well as
individual frames were divided into 16 overlapping subframes, and the
procedure was applied to each subframe separately.

Next, a {\em master list} of stars was compiled for each cluster based on
the best averaged frame of that cluster (henceforth {\em master frame}).
To that end the master frame was divided into 16 subframes, and the
DAOPHOT/ALLSTAR package \citep{ste87} was run for each subframe, assuming
a Moffat function with linear spatial variability to characterize the PSF.
Because of crowding the compilation proceeded in an iterative way, gradually
decreasing the detection threshold. We took care to avoid artificial
splitting of bright stars which can happen when an automatic procedure is used
to detect missed objects in subtracted images. During the final iteration
the residual images were inspected by eye to find previously undetected objects.

The listed stars were subsequently identified in the remaining averaged
frames of a given cluster, and profile photometry as well as PSF modeling
was performed for those frames with the ALLSTAR parameter {\em REDET} set to 1,
i.e. allowing for the re-determination of coordinates.
\section {Proper motions}
To derive the PMs we employed the same method as \citet{zlo11}. For convenience,
we briefly review the method in this  section.
\subsection{Measurements}
\label{ssect:measurements}
In each averaged frame of a given cluster the position of each star from the
master list was determined with respect to nearby cluster members using a
procedure similar to that described by \citet{and06}. For the first
guess we defined cluster members as objects located on the main sequence, red
giant branch and horizontal branch of the CMD of the cluster. To select them 
from the master list a $V/(V-I)$ or $V/(B-V)$ CMD was constructed, 
composed of stars with $V<21.0$ mag
and good photometry. The photometric quality was evaluated based on fit
parameters {\em CHI} and {\em SHARP} returned by ALLSTAR: stars with
$0.02<CHI<1.00$ and $-0.3<SHARP<0.3$ were only included.

Next, to each star a set of {\em grid objects} was assigned, composed of
cluster members located in an $\sim$80$\times80$ arcsec square centered on
that star. The grid objects were used to define the local geometrical
transformation between the master frame and the averaged frame being processed.
Typically, there were $\sim$200 of grid objects per star. For the transformation
function a two-dimensional 3rd order Chebyshev polynomial was chosen, whose
parameters were calculated with the help of
IRAF\footnote{IRAF is distributed by the National Optical Astronomy
Observatories, which are operated by the AURA, Inc., under cooperative
agreement with the NSF.} tasks immatch.geomap and immatch.geoxytran.

The $(X_0, Y_0)$ coordinates of the star on the master frame
were transformed into the expected coordinates $(X_C, Y_C)$ on the averaged
frame, and relative motions $dX = X_C-X_0$ and $dY = Y_C-Y_0$ were calculated.
Finally, using relative motions from all suitable averaged frames, the PMs
of the star $\mu_X$ and $\mu_Y$ were obtained from linear least-square
fits to $dX$ and $dY$ as functions of time.
The fitting was only attempted for objects with positions determined in at
least four epochs spanning at least four years. The confidence level was set
to 99\%, i.e. all stars for which the significance of the fit was smaller
than 99\% were rejected. Grid stars for which reliable
PMs had been obtained were then shifted to positions corresponding to the
epoch at which the reference frame was taken. Transformations and fits were
then repeated. The final total PMs were calculated from the
standard formula $\mu=(\mu_X^2+\mu_Y^2)^{1/2}$.

The catalog of the derived PMs, whose first few lines are shown in Table
\ref{tab:catalog}, is available online at http://case.camk.edu.pl. Altogether
it contains data for 13036, 12654, 10961, 22544, 11781 and 16394 stars 
in M4, M12, M22, NGC~3201, NGC~6362 and NGC~6752, respectively
Equatorial coordinates of the stars were obtained using frame astrometric
solutions based on a set of stars with $V < 17$ mag selected from the UCAC3 catalog
\citep{zac10}. The average residual of the adopted solution varied between
0.15 arcsec for NGC~6362 and 0.21 arcsec for M4.

\subsection {Error estimates}
 \label{ssect:error}
 
In the overlapping part of the two M4 fields there were 10385 stars with measured 
PMs (in the case of NGC~3201 fields there were 17693 such stars). For 66 per cent
of them the difference between PMs measured in each field separately
was smaller than 0.16 (0.08) mas/yr, which can be adopted as a robust
estimate of the average error of the PM determination. For 95 per cent
of the stars the difference was smaller than 0.75~(0.37)~mas/yr.
The median differences were nearly the same in all magnitude ranges.

The formal error $\sigma_\mu$ of the total proper motion was derived
for each star from the least-square fit described in Sect.
\ref{ssect:measurements}. For stars with $V\approx19.0$ mag the median
value of $\sigma_\mu$ varied from 0.29 to 0.67 mas/yr (observed
in NGC~3201 and M4, respectively). Not surprisingly, these limits were
increasing for fainter stars, and at $V\approx20$ mag they reached 0.46 and
1.62 mas/yr (observed, respectively, in NGC~3201 and NGC~6362).
An increasing scatter in $\sigma_\mu$ was also observed in all clusters
for $V < 14.5$ mag. This is due to the saturation of bright stars on some
images, resulting in fewer averaged frames available for the PM measurement.
In general, we may say that for $V<17.0$ mag the formal PM errors result
mainly from systematic effects, while at fainter magnitudes their main
source is photon noise.
\begin{landscape}
\setlength{\tabcolsep}{4pt}
\begin{table*}
\caption{
First few lines of the PM catalog for M4. Columns: (1) 
star ID (starting with 1 and 2 for fields F1 and F2, respectively); (2) \& (3) 
equatorial coordinates $(\alpha,\delta)_{2000.0}$ for epochs 1995.41 and 2008.45 
for F1 and F2, 
respectively; (4) \& (5) XY pixel coordinates on reference frames; (6)--(9) PMs and 
their errors; (10) number of epochs used;
(11) temporal baseline; (12) cluster membership (for explanation see Sec. 3.4);
(13) $V$-magnitude; (14) error of $V$; (15) $B-V$ ; (16) error of $B-V$. An entry
of 9.999 means that the corresponding quantity has not been measured. 
The whole catalog is available online at case.camk.edu.pl. 
\label{tab:catalog}}
\begin{tabular}{@{}cccccccccccccccc}
\hline
ID & $\alpha$ & $\delta$ & $X$ & $Y$ & $\mu_{\alpha}cos\delta$ & $\sigma_{\mu_{\alpha}cos\delta}$ & $\mu_{\delta}$ & $\sigma_{\mu_{\delta}}$ & N & dT & $mem$ & $V$ & $\sigma_{V}$ & $B-V$ & $\sigma_{B-V}$ \\
(1) & (2) & (3) & (4) & (5) & (6) & (7) & (8) & (9) & (10) & (11) & (12) & (13) & (14) & (15) & (16) \\
  {\scriptsize[\#]} 
& {\scriptsize[$^\circ$]} 
& {\scriptsize[$^\circ$]} 
& {\scriptsize[pixel]} 
& {\scriptsize[pixel]} 
& {\scriptsize[mas/yr]} 
& {\scriptsize[mas/yr]} 
& {\scriptsize[mas/yr]} 
& {\scriptsize[mas/yr]} 
& {\scriptsize[\#]}
& {\scriptsize[yr]} 
& {\scriptsize[-]} 
& {\scriptsize[mag]} 
& {\scriptsize[mag]} 
& {\scriptsize[mag]} 
& {\scriptsize[mag]} \\
\hline
1110412 &245.975229& -26.575613&  354.062&  20.026&   -0.19&   0.31&  -1.09&  0.87&  6&  5.086& 2& 20.467&  0.027&  1.332&  0.026\\
1110303 &245.973278& -26.589989&  155.339&  43.471&   -1.40&   0.82&   0.19&  0.64&  8& 11.076& 2& 19.385&  0.008&  1.069&  0.012\\
1110429 &245.972717& -26.582292&  261.684&  50.800&    0.12&   0.97&  -0.14&  0.38& 19& 14.081& 2& 20.586&  0.028&  1.507&  0.023\\
1110294 &245.971364& -26.570567&  423.667&  68.152&   10.86&   1.08&  12.26&  0.39& 26& 14.081& 0& 19.332&  0.010&  1.337&  0.018\\
1110124 &245.969529& -26.593671&  104.340&  89.706&   -0.12&   0.24&  -0.31&  0.26&  8& 11.076& 2& 17.842&  0.004&  0.847&  0.009\\
1110150 &245.969449& -26.574499&  369.270&  91.668&    8.35&   0.31&  14.72&  0.12&  8& 11.076& 0& 18.139&  0.004&  1.135&  0.010\\
1110197 &245.969333& -26.582213&  262.658&  92.719&   -0.05&   0.36&  -0.04&  0.35&  8& 11.076& 2& 18.527&  0.004&  0.909&  0.010\\
1110454 &245.969287& -26.587998&  182.719&  93.000&    0.47&   1.14&   1.07&  1.03& 27& 14.081& 2& 20.851&  0.033&  1.491&  0.034\\
1116026 &245.969344& -26.564876&  502.233&  93.448&    1.11&   1.08&   1.73&  1.67& 36& 14.081& 0& 21.059&  0.047&  1.515&  0.057\\
1110450 &245.969205& -26.566927&  473.888&  95.064&   -3.33&   1.13&  -1.86&  1.26& 11& 11.072& 0& 20.836&  0.030&  1.423&  0.025\\
1110364 &245.968586& -26.573083&  388.804& 102.426&    0.62&   0.67&  -0.65&  0.80&  8& 11.076& 2& 19.994&  0.013&  1.294&  0.016\\
1110460 &245.967451& -26.596180&   69.589& 115.308&    0.33&   1.50&   0.10&  0.74& 25& 14.081& 2& 20.936&  0.033&  1.512&  0.029\\
1110230 &245.966177& -26.579653&  297.929& 131.927&    0.52&   0.31&   0.31&  0.57&  8& 11.076& 2& 18.811&  0.006&  0.987&  0.011\\
1116023 &245.964502& -26.581214&  276.299& 152.590&   -0.48&   0.46&  -0.74&  0.30&  8& 11.076& 2& 18.482&  0.010&  0.943&  0.020\\
1110417 &245.963061& -26.564212&  511.200& 171.276&   -0.91&   0.87&   1.34&  1.22& 12& 11.072& 2& 20.497&  0.023&  1.543&  0.025\\
1110240 &245.962554& -26.574815&  364.663& 177.022&    6.59&   0.77&  19.90&  0.68&  8& 11.076& 0& 18.922&  0.026&  1.410&  0.039\\
1110411 &245.962965& -26.565007&  500.205& 172.428&    7.49&   0.60&  13.13&  0.78& 12& 11.072& 0& 20.457&  0.013&  0.839&  0.022\\
1110355 &245.962111& -26.572929&  390.700& 182.599&   -1.06&   0.62&  -0.00&  0.34&  8& 11.072& 2& 19.894&  0.012&  1.227&  0.017\\
1116024 &245.961945& -26.573605&  381.363& 184.615&   -1.34&   0.84&   0.52&  0.72&  8& 11.076& 2& 19.996&  0.015&  1.292&  0.024\\
1110389 &245.962454& -26.598838&   32.689& 177.032&    0.72&   0.88&   1.00&  0.84& 25& 14.081& 2& 20.188&  0.019&  1.295&  0.020\\
1110327 &245.960497& -26.569715&  435.069& 202.742&   -0.07&   0.41&   0.94&  0.77& 10& 11.076& 2& 19.614&  0.008&  1.148&  0.015\\
1110127 &245.959694& -26.598007&   44.089& 211.241&   -0.44&   0.22&  -0.37&  0.42&  7&  5.087& 2& 17.899&  0.004&  0.833&  0.013\\
1110288 &245.958568& -26.571603&  408.911& 226.518&   -0.96&   0.42&   1.28&  0.21&  8& 11.076& 2& 19.279&  0.008&  1.068&  0.012\\
1110299 &245.956955& -26.579173&  304.249& 246.098&    9.64&   0.37&  15.23&  0.40&  8& 11.076& 0& 19.362&  0.009&  1.119&  0.013\\
1110298 &245.956527& -26.574304&  371.513& 251.637&   -0.91&   0.44&  -0.38&  0.69&  8& 11.072& 2& 19.356&  0.009&  1.123&  0.014\\
1110418 &245.955256& -26.568132&  456.756& 267.677&   -0.12&   1.11&   1.26&  1.04& 35& 14.081& 2& 20.503&  0.030&  1.516&  0.025\\
1110238 &245.954365& -26.578801&  309.290& 278.163&   -1.15&   0.61&   0.49&  0.47&  8& 11.076& 2& 18.904&  0.006&  1.006&  0.011\\
1110198 &245.953445& -26.594167&   96.936& 288.752&    1.25&   0.35&   0.09&  0.33&  8& 11.076& 2& 18.531&  0.005&  0.905&  0.014\\
\hline
\end{tabular}
\end{table*}
\end{landscape}

\subsection{Completeness}
\label{ssect:completness}
We defined the completeness of our survey as the ratio of the
number of stars for which the PM was successfully measured to
the number of stars for which the PM measurement was attempted.
We assessed it as a function of $V$-magnitude and radial
distance from the center of the cluster $r$. In most clusters the
completeness exceeds $\sim$70 per cent for $13 < V < 17$ mag and
drops to $\sim$ 25 per cent at $V = 20.0$ mag. The exception is M22,
where it is limited to $\sim$50 per cent, which is understandable
given heavy crowding of the field. As expected, the completeness
increases with $r$, flattening at about 4 arcmin. The drop
observed at small $r$ is due to crowding, and the similar
drop at large distances - due to the fact that for more distant
stars only a few epochs were often available. No attempt was made
to estimate the completeness of master lists. 
As it is for other ground based studies, this parameter is a strong  
function of the distance from the cluster center.
\subsection{Cluster membership}
\label{ssect:membership}
\begin{figure}
\includegraphics[width=0.75\textwidth, bb= 159 1249 4702 5910,clip]{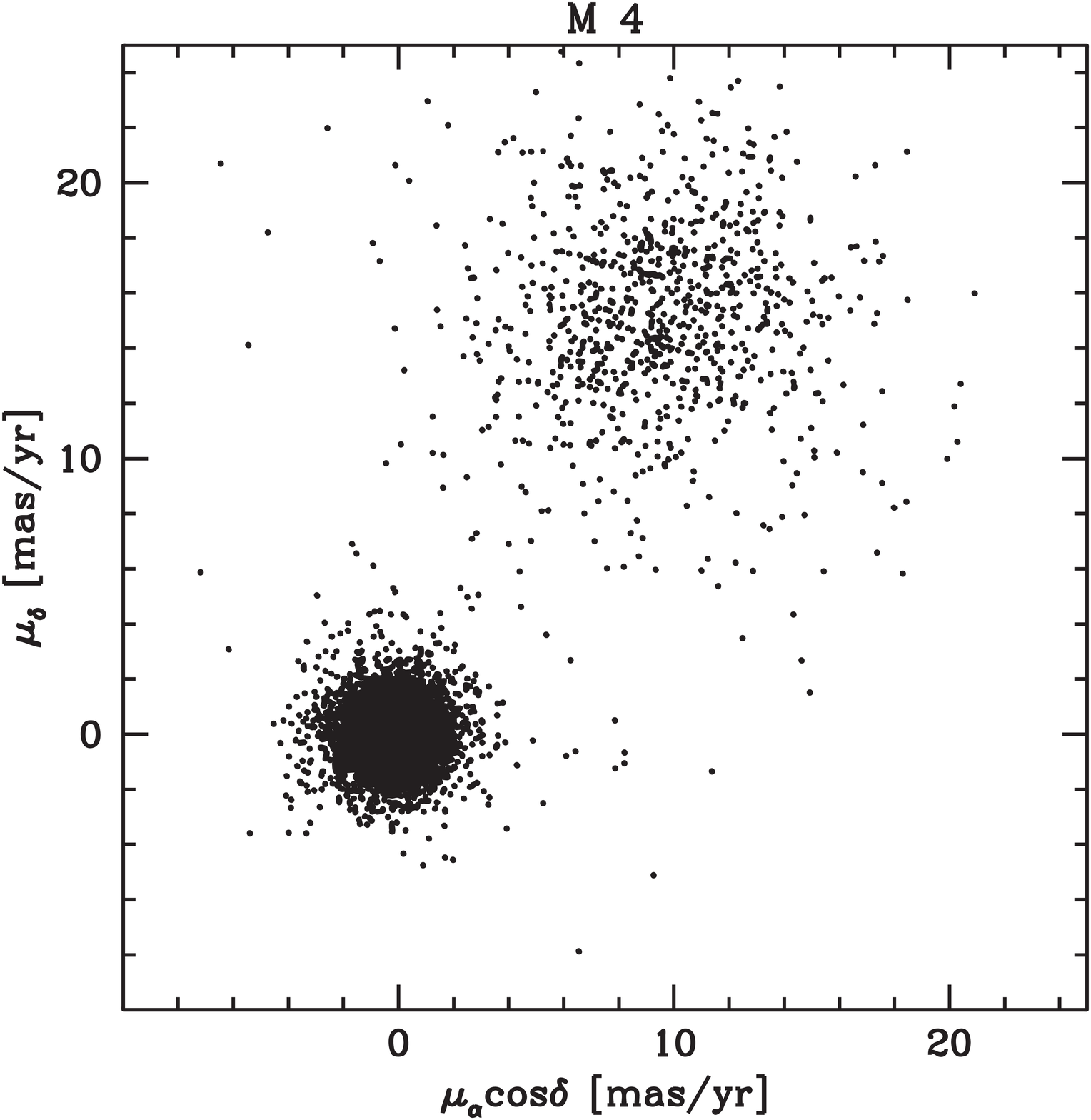}
 \caption {Results for M4. Vector point diagram for 13038 stars with
           measured relative proper motions.
           \label{fig:M4_VPD}
          }
\end{figure}

\begin{figure}
\includegraphics[width=0.75\textwidth, bb= 159 1249 4702 5910,clip]{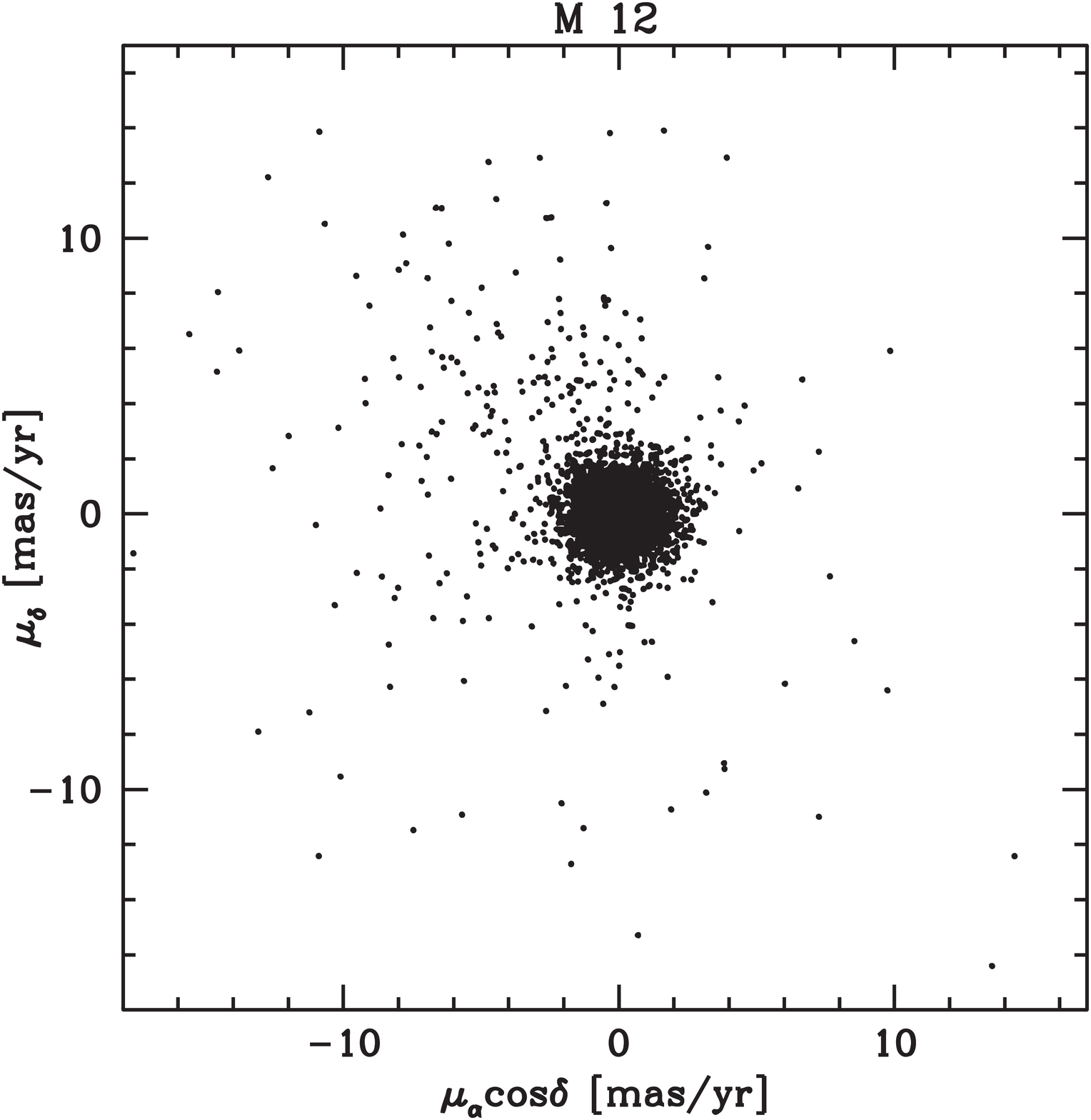}
 \caption {Results for M12. Vector point diagram for 12654 stars with
           measured relative proper motions.
           \label{fig:M12_VPD}
          }
\end{figure}

\begin{figure}
\includegraphics[width=0.75\textwidth, bb= 159 1249 4702 5910,clip]{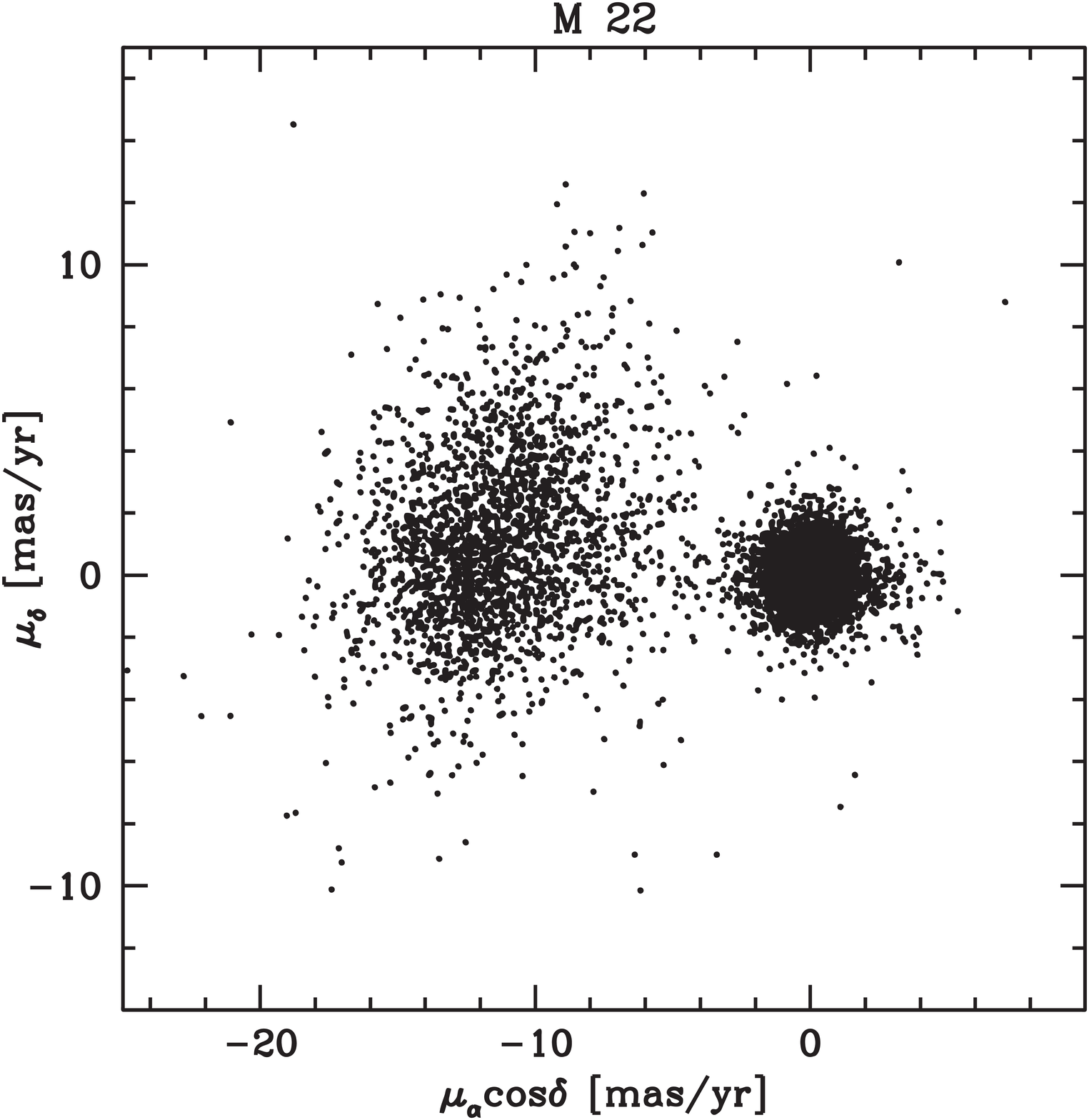}
 \caption {Results for M22. Vector point diagram for 10961 stars with
           measured relative proper motions.
           \label{fig:M22_VPD}
          }
\end{figure}

\begin{figure}
\includegraphics[width=0.75\textwidth, bb= 159 1249 4702 5910,clip]{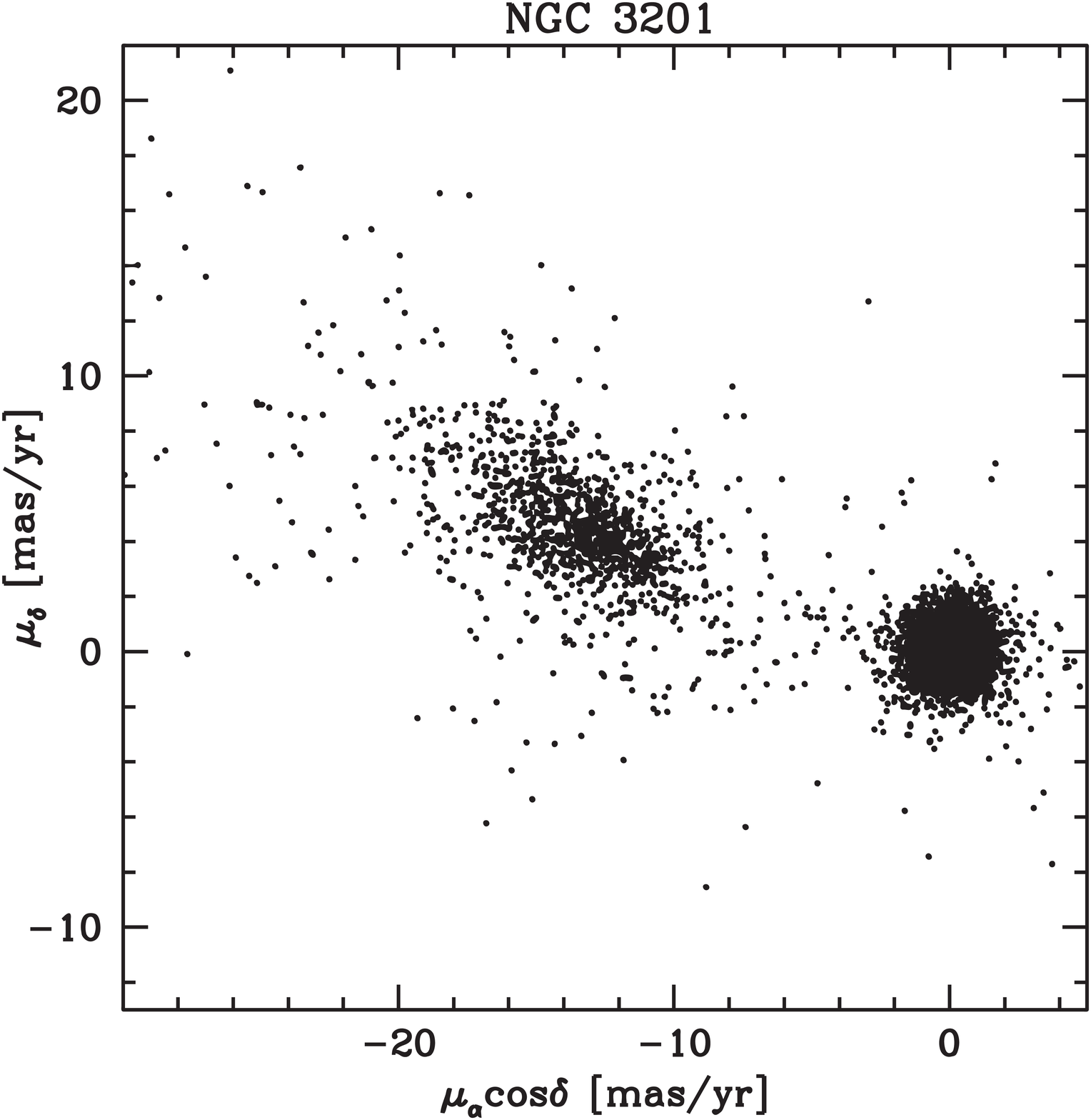}
 \caption {Results for NGC~3201. Vector point diagram for 22544 stars with
           measured relative proper motions.
           \label{fig:N3201_VPD}
          }
\end{figure}

\begin{figure} 
\includegraphics[width=0.75\textwidth, bb= 159 1249 4702 5910,clip]{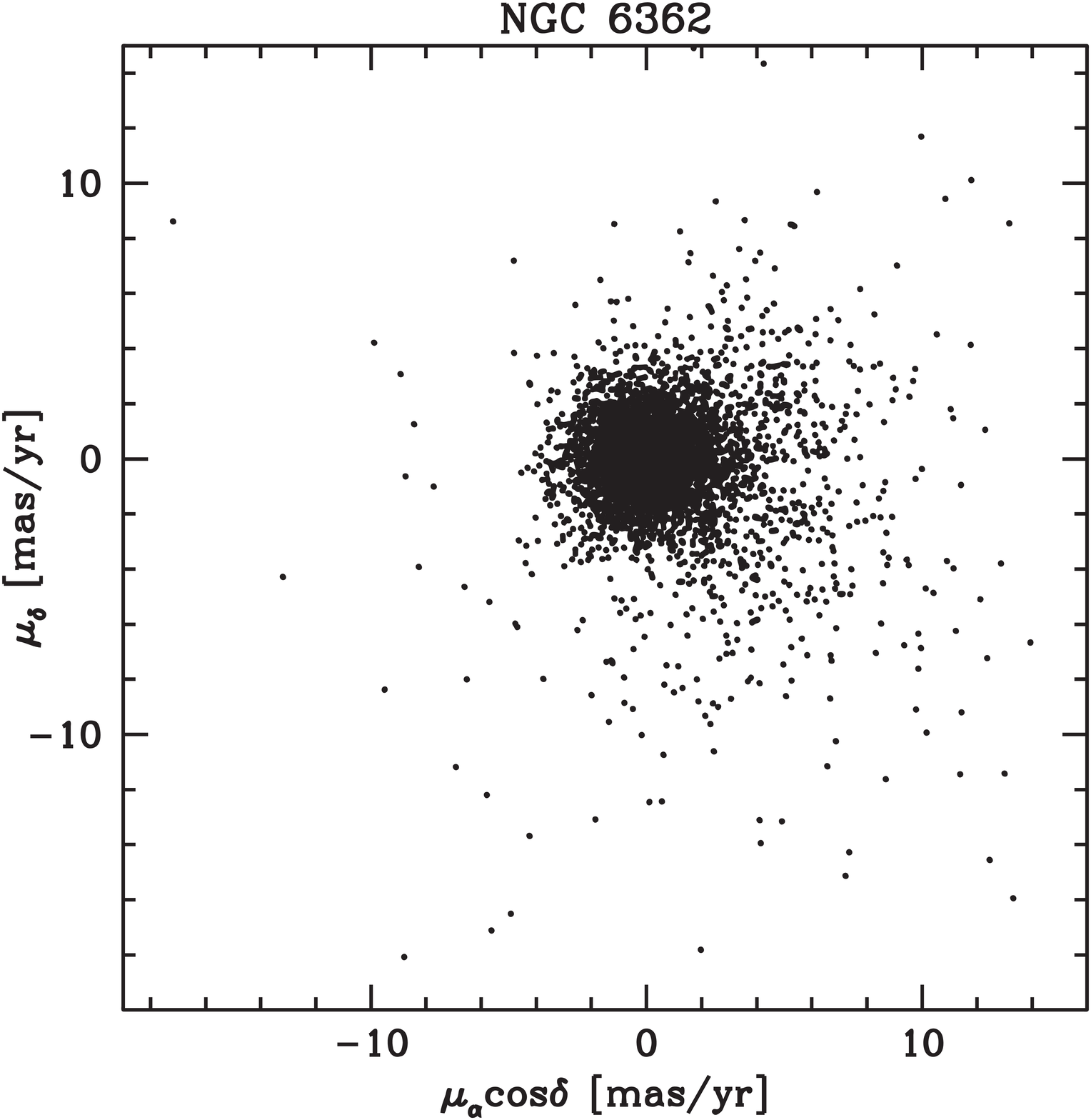}
 \caption {Results for NGC~6362. Vector point diagram for 11788 stars with
           measured relative proper motions.
           \label{fig:N6362_VPD}
          }
\end{figure}

\begin{figure}
 \includegraphics[width=0.788\textwidth]{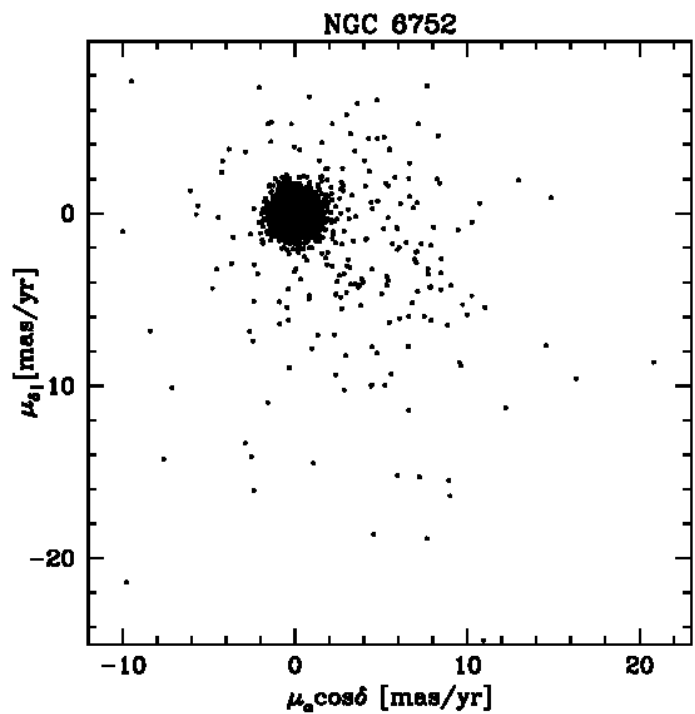}
 \caption {Results for NGC~6752. Vector point diagram for 12986 stars with
           measured proper motion.
           \label{fig:N6752_VPD}
          }
\end{figure}

\begin{figure*}
\includegraphics[width=1.00\textwidth]{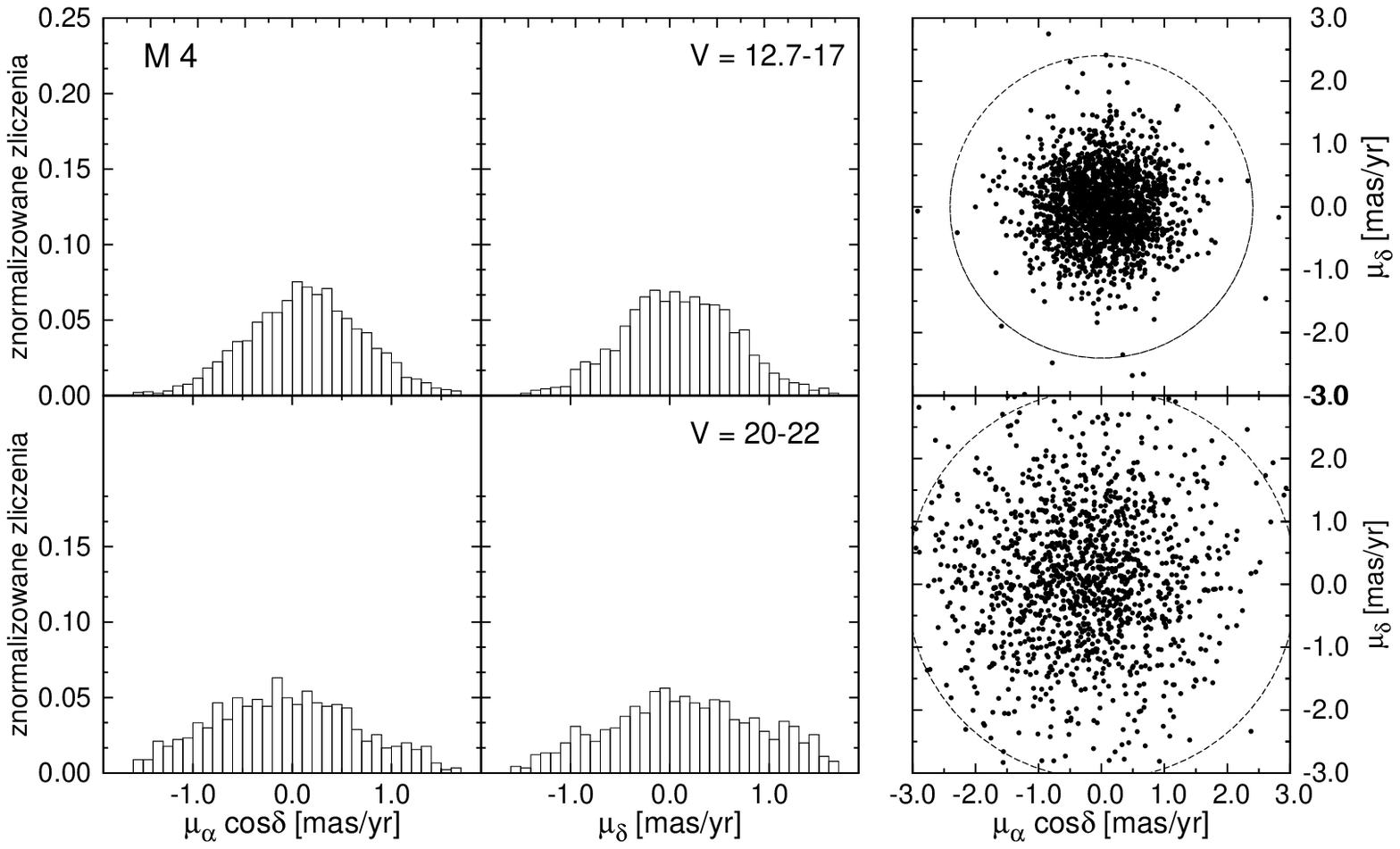}
 \caption {Results for M4. Distributions of $\mu_\alpha \cos\delta$
           and $\mu_\delta$ in bins $V=12.7-17$ and $V=20-22$ mag
           (left), and the corresponding vector point diagrams (right).
           The radii of the circles in the right panel are equal to $3S$
           with $S$ = 0.80 and 1.03 mas~yr$^{-1}$ in upper and lower
           diagram, respectively.
           \label{fig:M4_hist}
          }
\end{figure*}

\begin{figure*}
\includegraphics[width=1.00\textwidth]{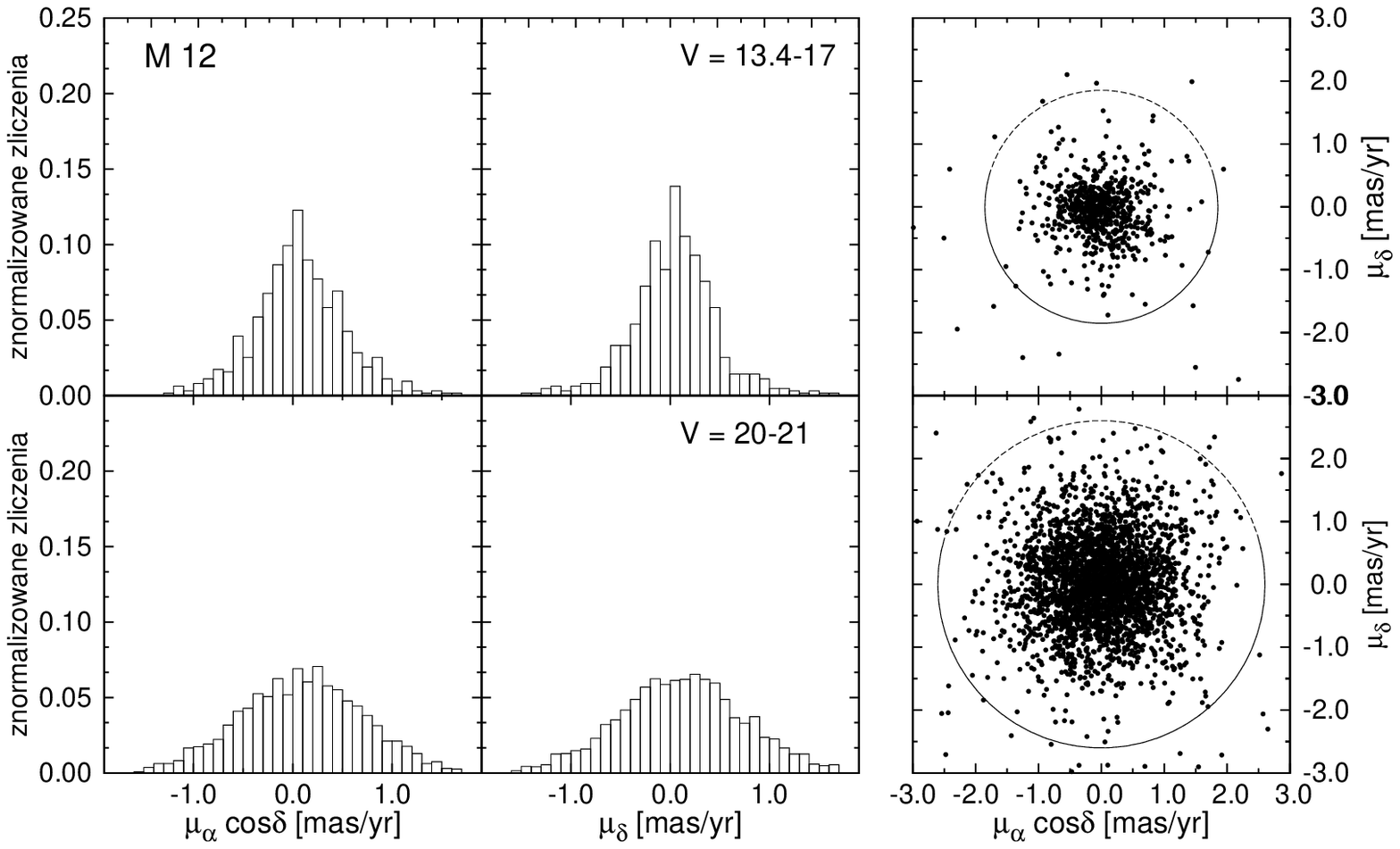}
 \caption {Results for M12. Distributions of $\mu_\alpha \cos\delta$
           and $\mu_\delta$ in bins $V=13.4-17$ and $V=20-21$ mag
           (left), and the corresponding vector point diagrams (right).
           The radii of the circles in the right panel are equal to $3S$
           with $S$ = 0.62 and 0.87 mas~yr$^{-1}$ in upper and lower
           diagram, respectively.
           \label{fig:M12_hist}
          }
\end{figure*}

\begin{figure*}
\includegraphics[width=1.00\textwidth]{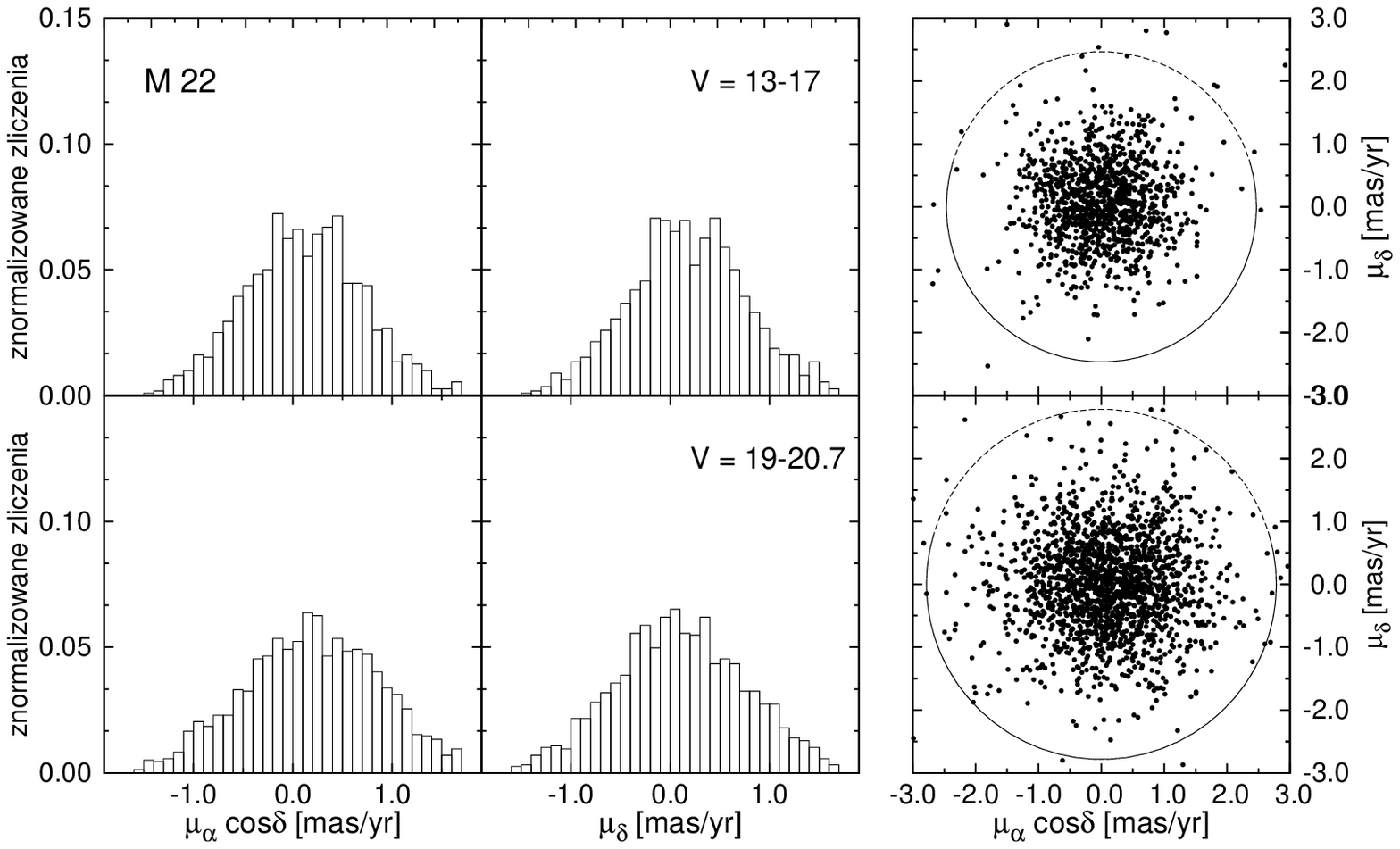}
 \caption {Results for M22. Distributions of $\mu_\alpha \cos\delta$
           and $\mu_\delta$ in bins $V=13.4-17$ and $V=20-21$ mag
           (left), and the corresponding vector point diagrams (right).
           The radii of the circles in the right panel are equal to $3S$
           with $S$ = 0.82 and 0.93 mas~yr$^{-1}$ in upper and lower
           diagram, respectively.
           \label{fig:M22_hist}
          }
\end{figure*}

\begin{figure*}
\includegraphics[width=1.00\textwidth]{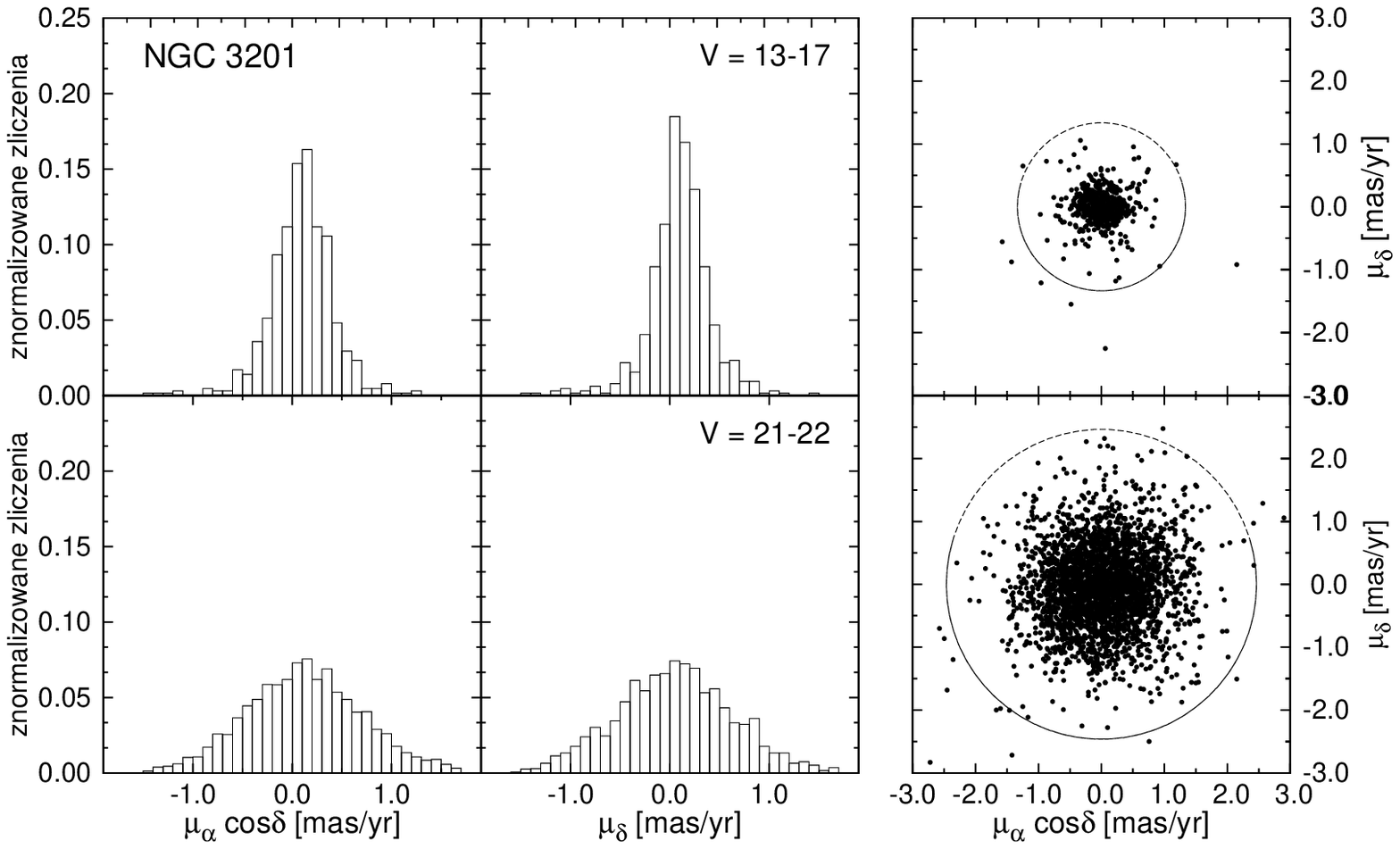}
 \caption {Results for NGC~3201. Distributions of $\mu_\alpha \cos\delta$
           and $\mu_\delta$ in bins $V=13-17$ and $V=20-21$ mag
           (left), and the corresponding vector point diagrams (right).
           The radii of the circles in the right panel are equal to $3S$
           with $S$ = 0.45 and 0.82 mas~yr$^{-1}$ in upper and lower
           diagram, respectively.
           \label{fig:N3201_hist}
          }
\end{figure*}

\begin{figure*}
\includegraphics[width=1.00\textwidth]{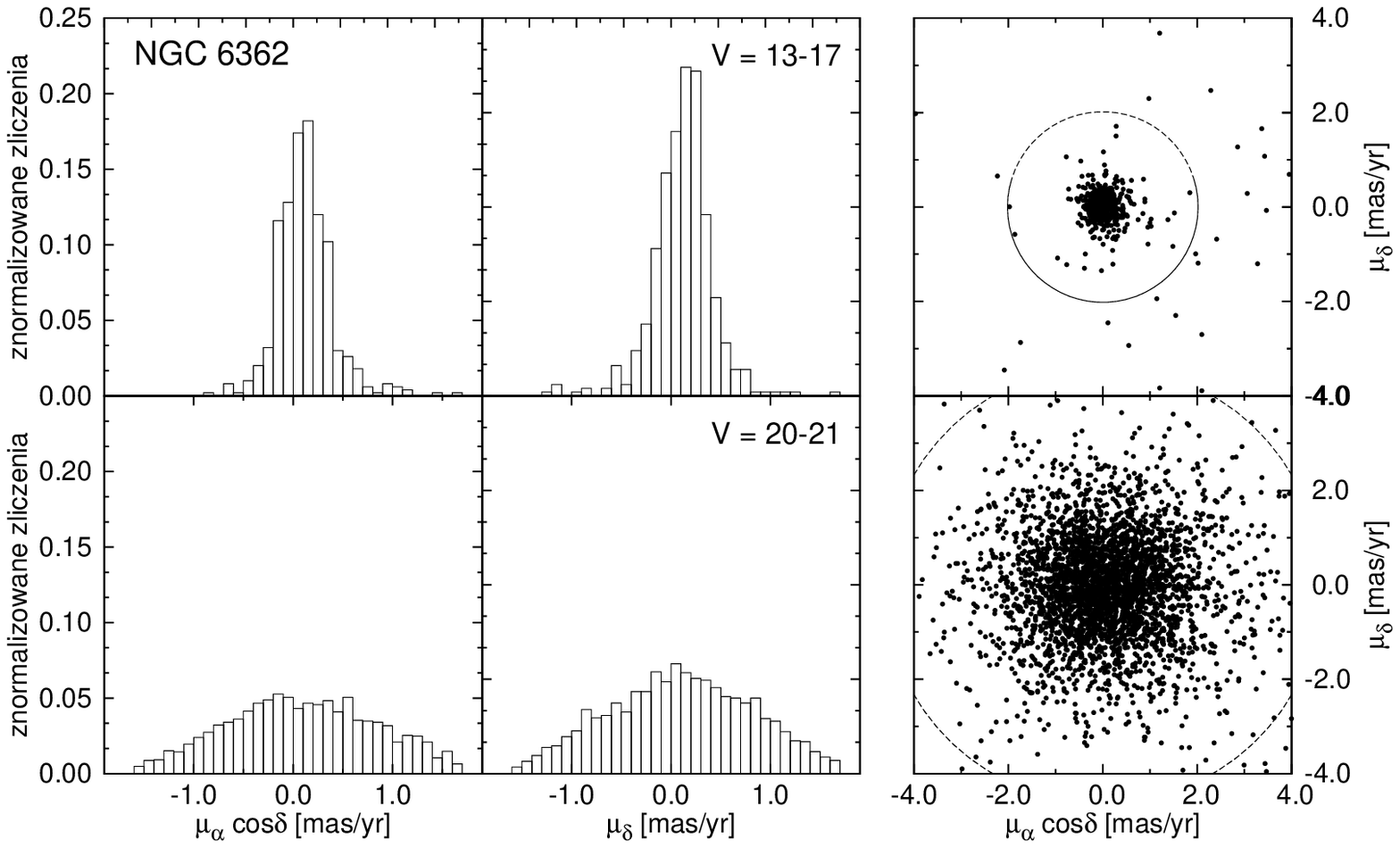}
 \caption {Results for NGC~6362. Distributions of $\mu_\alpha \cos\delta$
           and $\mu_\delta$ in bins $V=13-17$ and $V=20-21$ mag
           (left), and the corresponding vector point diagrams (right).
           The radii of the circles in the right panel are equal to $3S$
           with $S$ = 0.70 and 1.54 mas~yr$^{-1}$ in upper and lower
           diagram, respectively.
           \label{fig:N6362_hist}
          }
\end{figure*}

\begin{figure*}
\includegraphics[width=1.00\textwidth]{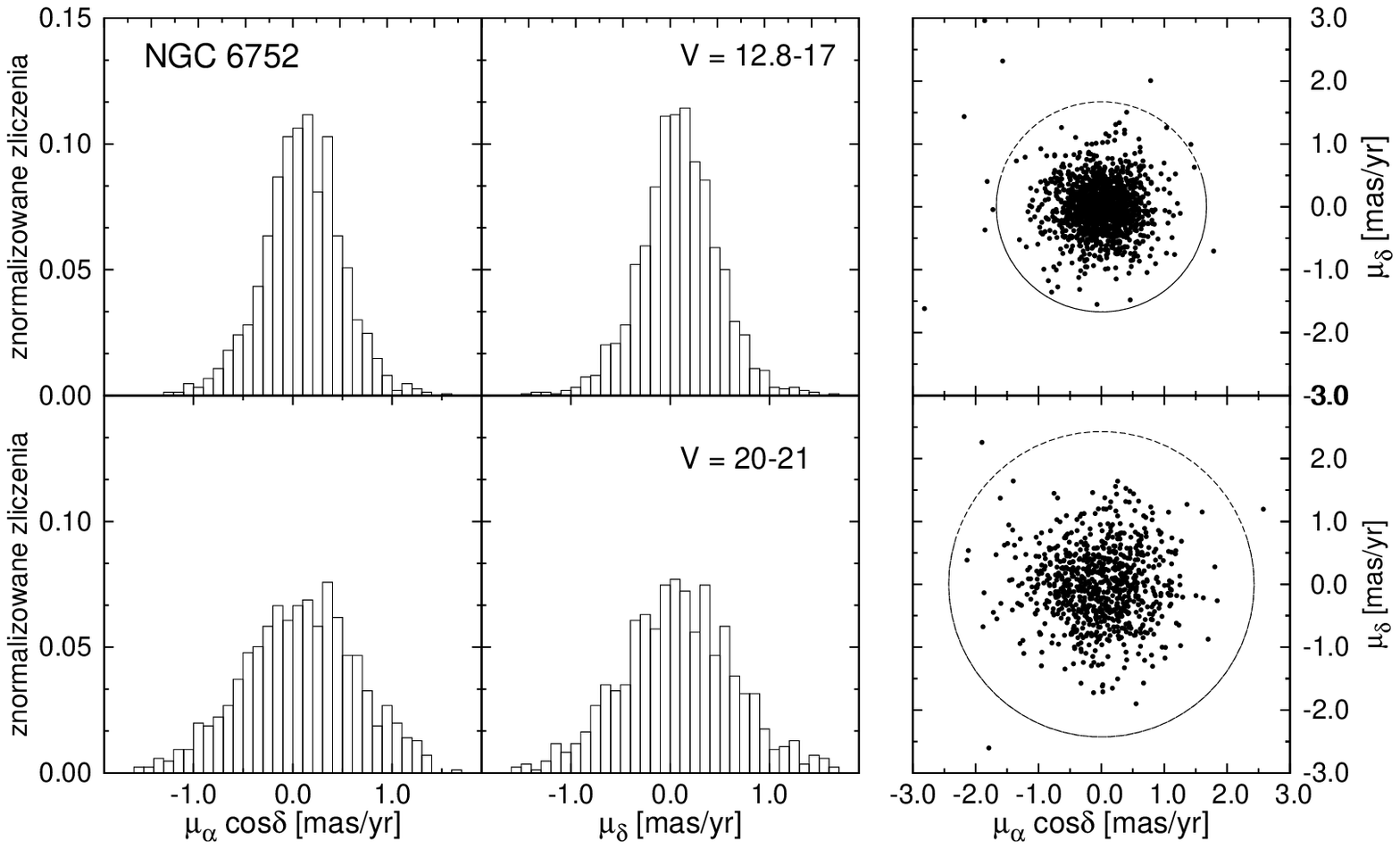}
 \caption {Results for NGC~6752. Distributions of $\mu_\alpha \cos\delta$
           and $\mu_\delta$ in bins $V=12.8-17$ and $V=20-21$ mag
           (left), and the corresponding vector point diagrams (right).
           The radii of the circles in the right panel are equal to $3S$
           with $S$ = 0.56 and 0.93 mas~yr$^{-1}$ in upper and lower
           diagram, respectively.
           \label{fig:N6752_hist}
          }
\end{figure*}

Vector point diagrams (VPD) showing stars with reliable PMs are
presented in Figs. \ref{fig:M4_VPD} -- \ref{fig:N6752_VPD}. In all
six cases cluster members are crowded around the (0,0) point,
and it is evident that field stars are a small fraction
of the total sample. 
For the clusters M4, M22 and NGC~3201 field stars form  
well defined clumps on the VPD that  barely overlap with the cluster
stars. For the remaining three clusters field stars do not show 
well defined clumps on the VPD. Instead they are spread over large
areas on the VPD and their distribution overlaps with cluster stars.
These properties of the VPDs and the small percentage of field stars in the
analyzed samples prevented us from estimating membership 
probabilities as it is usually done \citep[see e.g.][]{pla03}.
Instead, we assigned each star
to one of three membership classes ($mem$ = 0,1,2) based on its
location in the VPD and the uncertainty $\sigma_\mu$. 
Class 0 corresponds to likely field stars, class 1 to possible members 
and class 2  to likely members. The assignment
procedure consisted of the following steps:
\begin{enumerate}
\begin{table}
\caption{Number of stars belonging to membership classes 0, 1 and 2.
\label{tab:fraction}}
\begin{tabular}{@{}lrrr}
\hline
Cluster & 0 & 1 & 2\\
\hline
M4           & 2229  &  174 & 10633 \\
M12          & 1477  &   38 & 11139 \\
M22          & 2476  &   28 &  8457 \\
NGC~3201     & 3225  &  237 & 19082 \\
NGC~6362     & 1948  &  124 &  9709 \\
NGC~6752     & 1015  &  129 & 15250 \\
\hline
\end{tabular}
\end{table}
\item All stars were divided into magnitude bins containing at least
 $\sim$1000 objects.
\item In each bin mean values ($M_\alpha$, $M_\delta$) and standard
 deviations ($S_\alpha$, $S_\delta$) of the distributions of PM
 components in $\alpha$ and $\delta$ directions were found. The
 total standard deviation $S \equiv (S_\alpha^2 + S_\delta^2)^{1/2}$
 was calculated.
\item In each bin mean values ($ME_\alpha$, $ME_\delta$) and standard
 deviations ($SE_\alpha$, $SE_\delta$) of the distributions of the
 errors of PM components in $\alpha$ and $\delta$ directions were found.
 Total mean error $ME \equiv (ME\alpha^2+ME_\delta^2)^{1/2}$ and total
 standard deviation of the error $SE = (SE\alpha^2+SE_\delta^2)^{1/2}$
 were calculated.
\item Stars with $\mu > aS$, where $a$ was equal to 2.5 for M4 M22 and 
 NGC~3201, and to 2.0 for M12, NGC~6362 and NGC~6752, were classified as 
 nonmembers ($mem$ = 0). Stars with $\mu \leq aS$ and $\sigma_\mu > ME 
 + aSE$ were classified as possible members ($mem$ = 1), while those 
 with $\mu \leq aS$ and $\sigma_\mu\leq ME + aSE$ as genuine PM-members 
 of the cluster ($mem$ = 2).
\end{enumerate}
The histograms showing the
distributions of both PM components for the brightest and for the faintest
bin in $V$ are shown in Figs. \ref{fig:M4_hist} -- \ref{fig:N6752_hist}
together with the corresponding VPDs.
Table \ref{tab:fraction} lists the number of stars assigned
to membership classes 0, 1 and 2 for each cluster. 

\begin{figure*}
\includegraphics[width=7.5 cm, angle=270, bb= 245 75 562 590,clip]{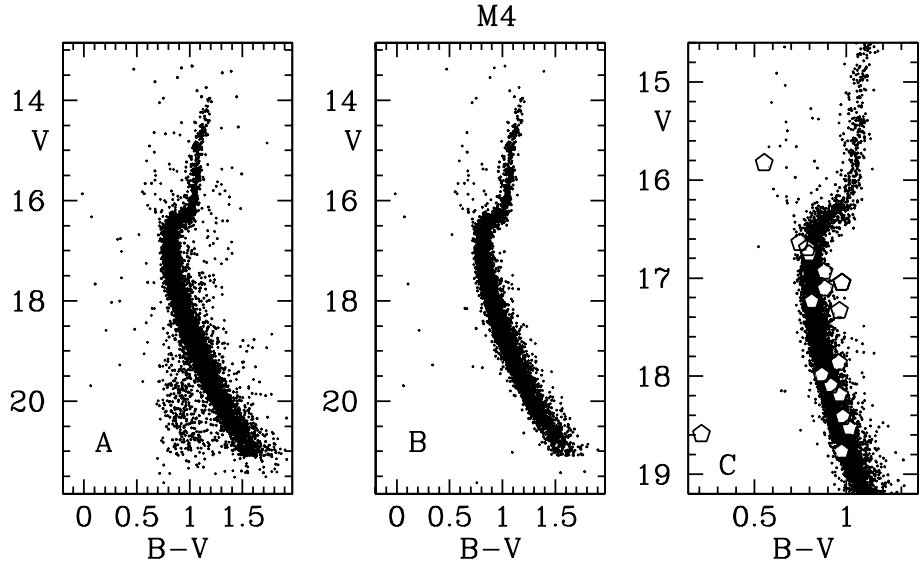}
 \caption {A: CMD of M4 showing all stars with measured relative proper motions and good 
           photometry. B: the same diagram showing likely cluster members 
           only, i.e. stars with $mem$ = 2. C: same as B but with marked 
           locations of known variable stars with $mem$ = 2.
           \label{fig:M4_cmd_clean}
          }
\end{figure*}

\begin{figure*}
\includegraphics[width=7.5 cm, angle=270, bb= 245 75 562 590,clip]{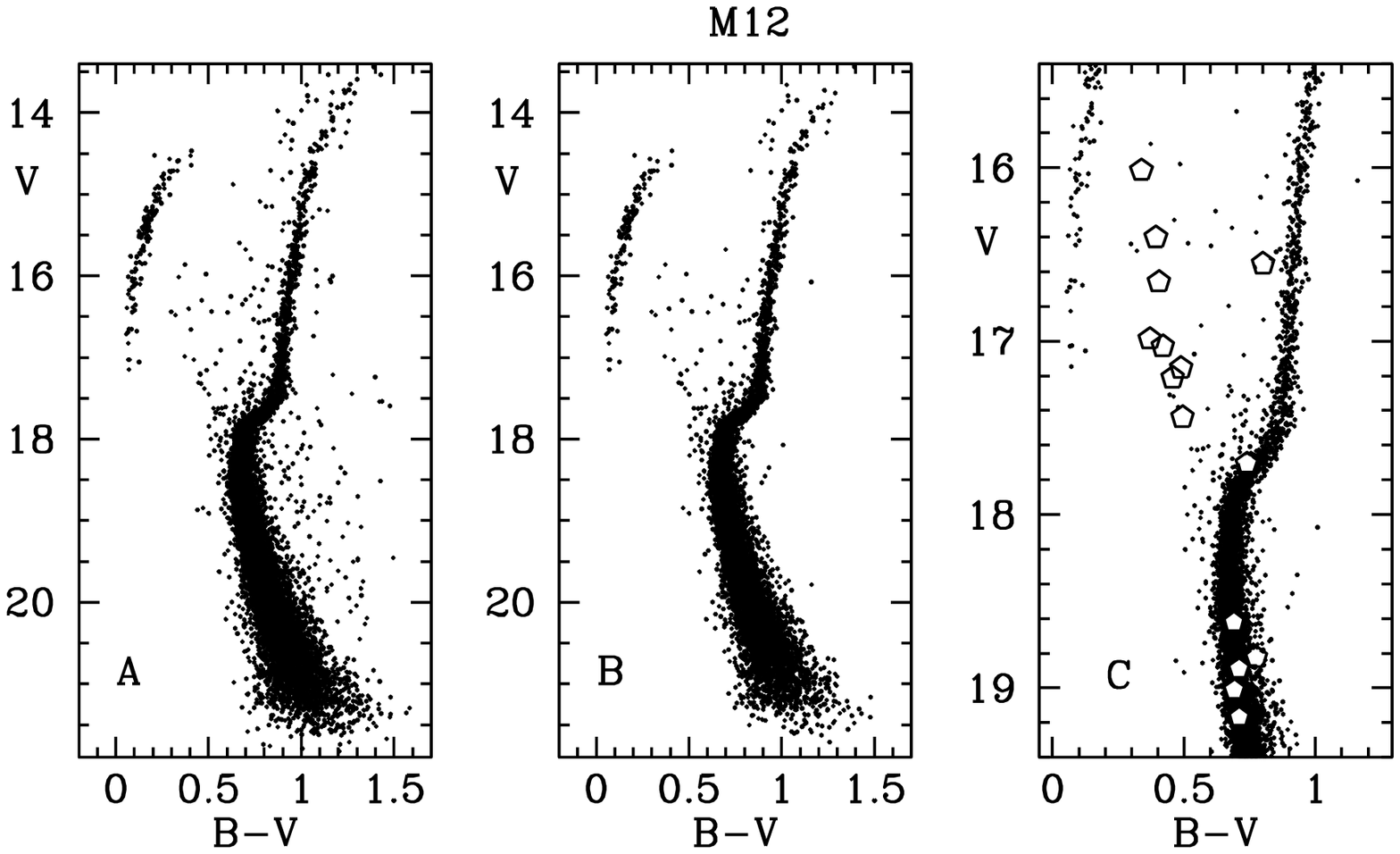}
 \caption {A: CMD of M12 showing all stars with measured relative proper motions. B:
           the same diagram showing cluster members only, i.e. stars
           with $mem$ = 2. C: same as B but with marked 
           locations of known variable stars with $mem$ = 2.
           \label{fig:M12_cmd_clean}
          }
\end{figure*}

\begin{figure*}
\includegraphics[width=7.5 cm, angle=270, bb= 245 75 562 590,clip]{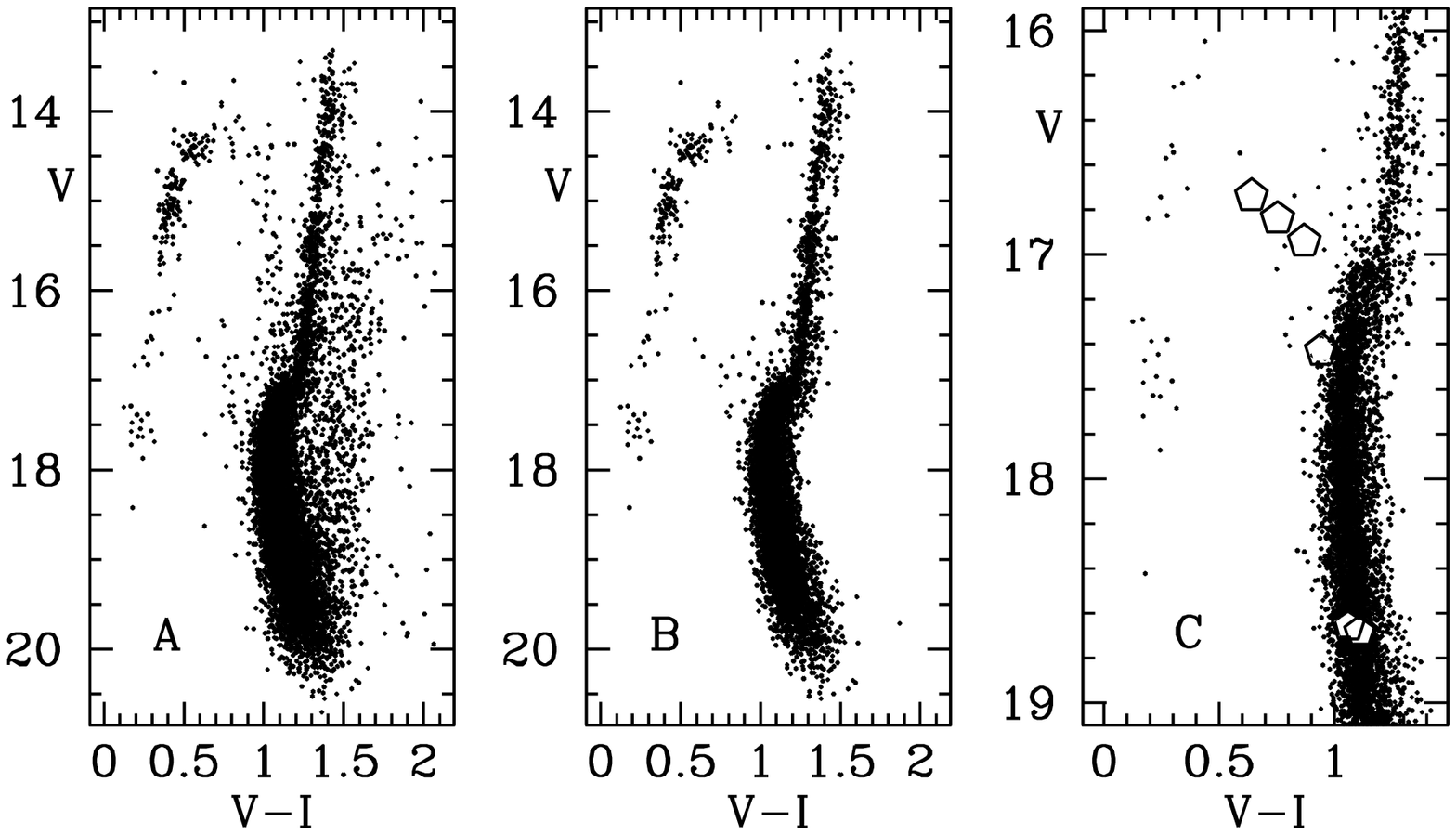}
 \caption {A: CMD of M22 showing all stars with measured relative proper motions. B:
           the same diagram showing cluster members only, i.e. stars
           with $mem$ = 2. C: same as B but with marked 
           locations of known variable stars with $mem$ = 2.
           \label{fig:M22_cmd_clean}
          }
\end{figure*}

\begin{figure*}
\includegraphics[width=7.5 cm, angle=270, bb= 245 75 562 590,clip]{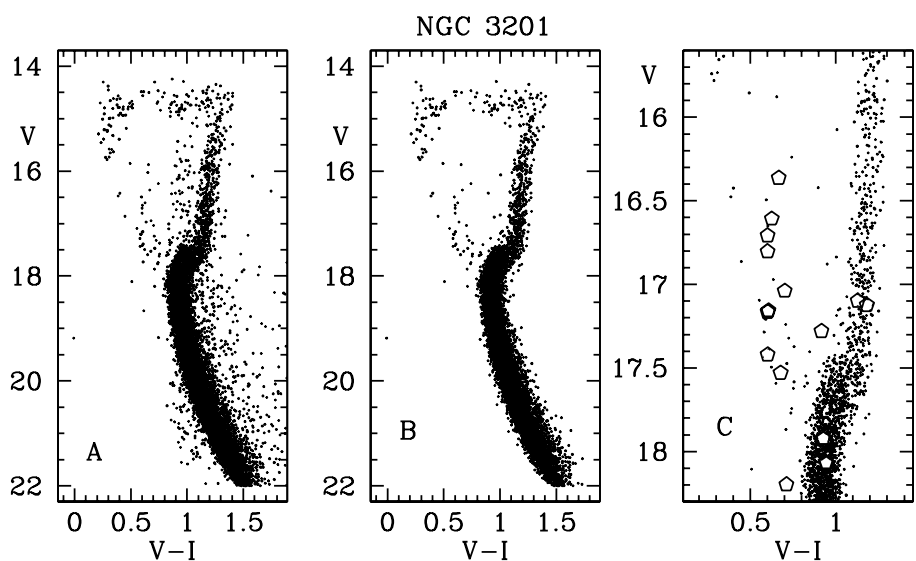}
 \caption {A: CMD of NGC~3201 showing all stars with measured relative proper motions. B:
           the same diagram showing cluster members only, i.e. stars
           with $mem$ = 2. C: same as B but with marked 
           locations of known variable stars with $mem$ = 2.
           In order to visualize the horizontal branch, bright stars with 
           poorer photometry were included in this plot.
           \label{fig:N3201_cmd_clean}
          }
\end{figure*}

\begin{figure*}
\includegraphics[width=7.5 cm, angle=270, bb= 245 75 562 590,clip]{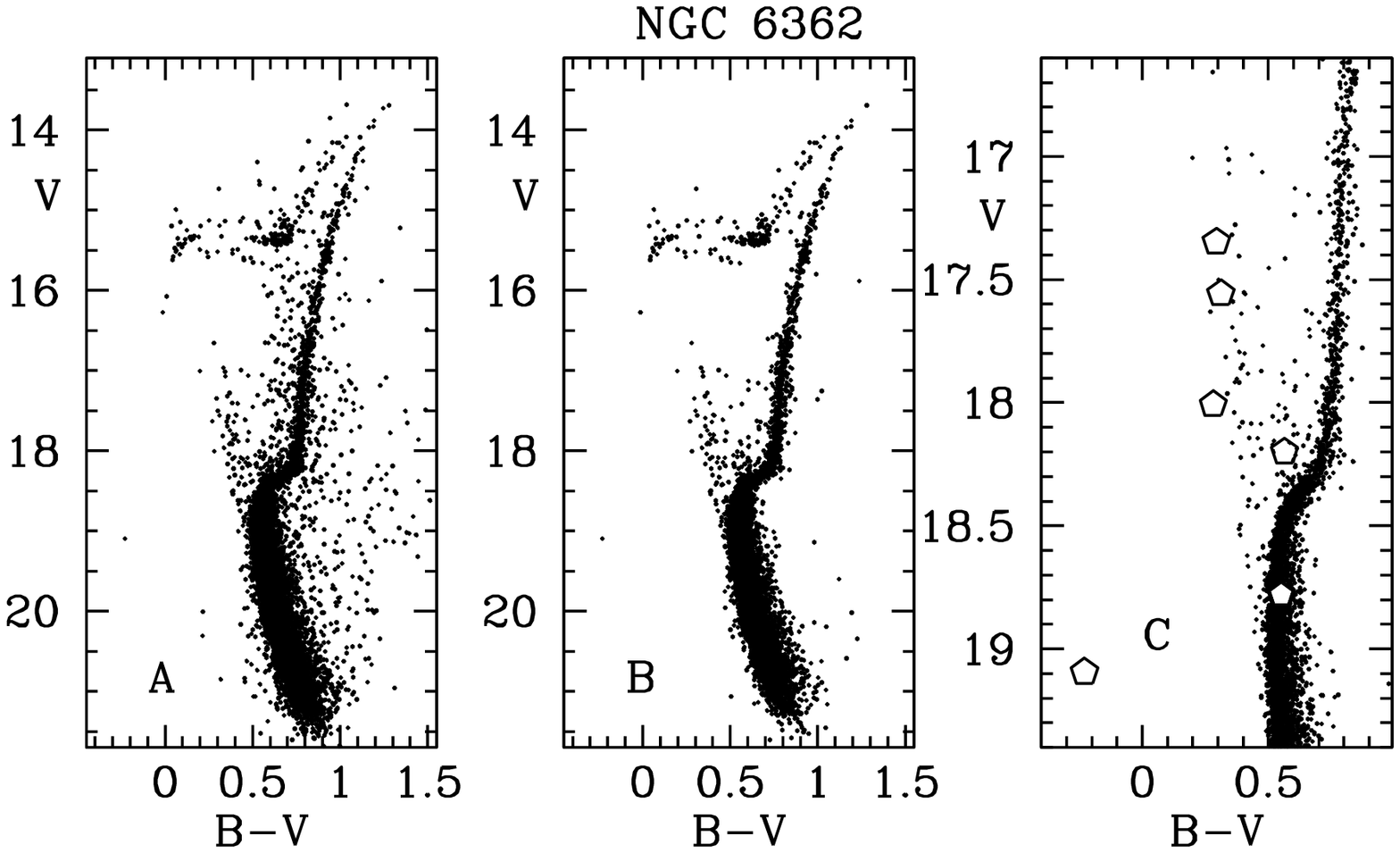}
 \caption {A: CMD of NGC~6362 showing all stars with measured relative proper motions. B:
           the same diagram showing cluster members only, i.e. stars
           with $mem$ = 2. C: same as B but with marked 
           locations of known variable stars with $mem$ = 2.
           \label{fig:N6362_cmd_clean}
          }
\end{figure*}

\begin{figure*}
\includegraphics[width=7.5cm, angle=270, bb= 245 75 562 590,clip]{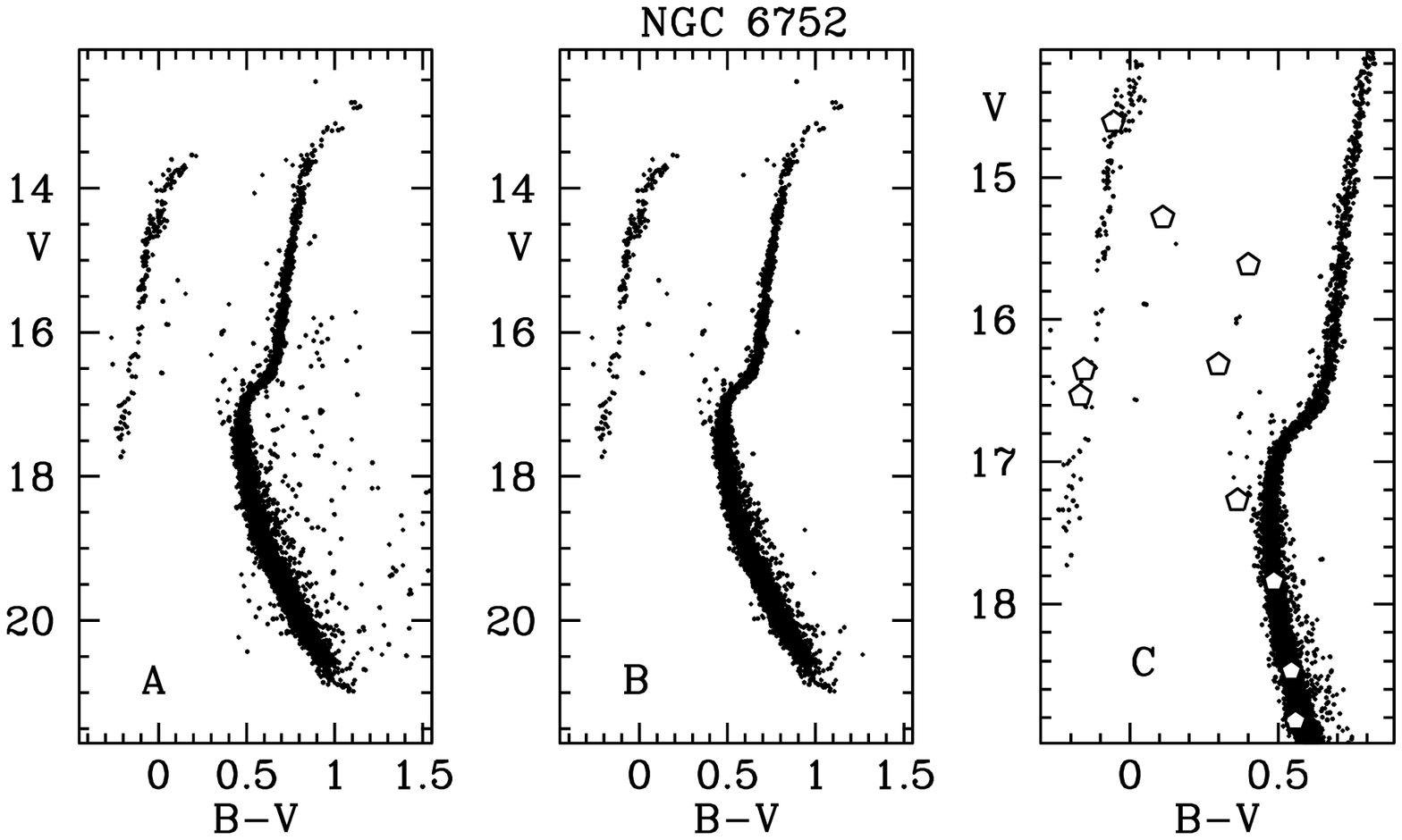}
 \caption {A: CMD of NGC~6752 showing all stars with measured relative proper motions. B:
           the same diagram showing cluster members only, i.e. stars
           with $mem$ = 2. C: same as B but with marked 
           locations of known variable stars with $mem$ = 2.
           \label{fig:N6752_cmd_clean}
          }
\end{figure*}

\Section{Color-magnitude diagrams}
Figures \ref{fig:M4_cmd_clean} -- \ref{fig:N6752_cmd_clean} show CMDs 
of the six analyzed cluster fields.
The photometry of M12, M22 and NGC~3201 was calibrated using 
local standards from \citet{ste00} catalog. For M4, NGC~6362 and
NGC~6752 the calibration was based on observations of \citet{land}
standards collected within CASE with instrumentation described in Sec. 2.  
Left panels of Figs. \ref{fig:M4_cmd_clean} -- \ref{fig:N6752_cmd_clean} 
include all stars with measured proper motion and with good photometry. 
Poor photometric measurements were filtered out by identifying stars with 
relatively large formal error of magnitude or color as compared with other 
stars of similar magnitude.   
Middle panels include only likely cluster members ($mem=2$). Right panels 
show a zoomed view of the turnoff region with the location of known 
variables identified. As for the middle panels, only stars with $mem=2$ are  
plotted. Lists of  variable stars were taken from published and 
unpublished CASE surveys.

Comments on individual clusters:\\

M4 -- In addition to about two dozen blue straggler stars,  
panel B of Fig. \ref{fig:M4_cmd_clean} shows about a dozen candidate
yellow straggler stars. Only one of the blue stragglers is a known variable.
Several likely cluster members are observed to the blue of the main sequence 
below turnoff region. One of those faint blue stars is the binary hot 
subdwarf V46 \citep{ktk,otoole}.\\

M12 -- This cluster shows about two dozen candidates for blue
and yellow stragglers (Fig. \ref{fig:M12_cmd_clean} B, C). Nine of them 
were identified as variables. A few likely members are observed to the red 
of the turnoff, and one is observed to the red of the subgiant branch. 
These are potentially interesting objects, deserving spectroscopic
follow-up. \\

M22 -- This cluster is located in front of the outer part of the 
Galactic bulge.
Comparison of panels A and B in Fig. \ref{fig:M22_cmd_clean} shows that 
field interlopers have been efficiently filtered out.
Relatively few candidates for blue stragglers, as compared with 
M4 and M12, are observed in panels B and C. The relatively large widths 
of the subgiant and giant branches results from a combination of differential reddening 
and chemical inhomogeneity of the cluster \citep{marino11}.    
One may also notice that the extended horizontal branch of 
M22 exhibits at least three gaps on the CMD. The pronounced
gap at $V=14.6$ mag has been  reported earlier by \citet{cho}
but  is more clearly seen in our data. Two other possible  gaps
are located at $V\approx 16$ mag and $V\approx 17$ mag. Most of the variables
reported so far by CASE for the central region of M22 turned out to 
be field stars \citep{pk03,kt01}.\\

NGC~3201 -- Like M22, this cluster suffers from differential reddening
causing a large width of the principal features of the CMD \citep{vonbraun}. 
As is apparent in panels B and C of Fig. \ref{fig:N3201_cmd_clean}, the cluster 
hosts a rich population of blue and yellow stragglers (note, however, that 
stars observed at $V<16$ mag and $B-V<0.4$ mag belong to the blue horizontal branch). 
Several blue stragglers are SX Phe type pulsating variables  \citep{mazur}. 
One hot subdwarf candidate is seen  at $V=19.2$ mag and $V-I=-0.01$ mag.  \\

NGC~6362 -- As we noted in Sect. \ref{ssect:membership}, cluster members 
overlap with field objects on the VPD (see Fig. \ref{fig:N6362_VPD}). 
Therefore, some likely interlopers can still remain in panels B and C of 
Fig. \ref{fig:N6362_cmd_clean}. 
Nonetheless, it is evident that the cluster hosts a rich population
of blue and yellow stragglers. The variable hot subdwarf identified by
\citet{mazur99} is a likely cluster member. In Fig. \ref{fig:N6362_cmd_clean}
this stars is located at $V=19.09$ mag and $B-V=-0.23$ mag. A pronounced asymptotic 
giant branch is clearly visible in the CMD. 
\\

NGC~6752 -- Several stars are located to the red of the subgiant branch and 
main-sequence in panel B of Fig. \ref{fig:N6752_cmd_clean}. The poor separation 
of cluster stars from field stars on the VPD (see Fig. \ref{fig:N6752_VPD}) 
suggests that these may be field interlopers not recognized by our simple 
membership criteria. Spectroscopic observations are needed to clarify 
status of these stars. On the other hand, a large fraction of the
variables identified 
by \citet{jka09} are likely cluster members.   


\section{Summary}
\label{sect:summary} 
Based on a uniform set of data collected between 1997 and 2008 we derived relative 
proper motions of stars in the fields of six globular clusters: M4, M12, M22, 
NGC~3201, 
NGC~6362 and NGC~6752. The measurements were made using a procedure similar 
to that described by 
\citet{and06}. Proper motions were successfully measured  for over 87000 stars and 
a membership class was assigned to each. The membership 
classes allow for an efficient elimination of field stars from color-magnitude 
diagrams of the clusters. A catalog of proper motions was compiled and is publicly 
available at case.camk.edu.pl. These relative proper motions aid in the
efficient selection of 
rare objects such as blue/yellow/red stragglers and stars from the asymptotic giant
branch. The data can be also used to assign membership status 
to variable stars, either already identified or to be detected by future surveys. 

\section*{Acknowledgments}
This research used the facilities of the Canadian Astronomy Data Centre
operated by the National Research Council of Canada with the support
of the Canadian Space Agency.

\clearpage
\clearpage
\clearpage
\clearpage
\clearpage
\clearpage
%
\clearpage
%
\clearpage
%
\end{document}